\documentclass{aa}

\usepackage{graphics}

\begin{document}

\renewcommand{\topfraction}{1.}
\renewcommand{\bottomfraction}{1.}
\renewcommand{\textfraction}{0.}

\thesaurus{06
           (08.01.2;
            08.03.4;
            08.06.1;
            08.09.2 FK~Comae;
            08.12.1;
            02.12.3)}

\title{Circumstellar emission and flares on FK~Comae Berenices}

\subtitle{Analysis and modelling of Balmer and \ion{He}{i} D3 line variations\thanks{based on observations obtained at the Isaac Newton Telescope with the ESA-MUSICOS spectrograph and at the Observatoire de Haute Provence with the Aur\'{e}lie spectrograph.}}

\author{J.M.~Oliveira \inst{1,2}
\and B.H.~Foing \inst{1,3}}

\offprints{J.M.~Oliveira, ESTEC address \\{\it Correspondence to}: joana@so.estec.esa.nl}

\institute{ESA Space Science Department, ESTEC/SCI-SO, P.O.~Box 299, NL-2200 AG Noordwijk, The Netherlands\\
           joana@so.estec.esa.nl, bfoing@estec.esa.nl
\and Centro de Astrof\'{\i}sica da Universidade do Porto, Rua das Estrelas s/n, P-4150 Porto, Portugal
\and Institut d'Astrophysique Spatiale, CNRS/Univ.~Paris~XI, B\^{a}t.~121, Campus d'Orsay, F-91405 Orsay Cedex, France}

\date{Received 27 August 1998 / Accepted 1 December 1998}

\titlerunning{Circumstellar emission and flares on FK Comae}
\authorrunning{Oliveira \& Foing}

 \maketitle

\begin{abstract}

We present results on spectroscopic observations of the fast-rotating active giant \object{FK Comae}, obtained mainly with the ESA-MUSICOS spectrograph at the Isaac Newton Telescope (INT) in 1996 and 1997 and also with the Aur\'{e}lie spectrograph at the Observatoire de Haute Provence (OHP) in 1997. The profiles analysed are those of the Balmer, H$\alpha$\ and H$\beta$, and \ion{He}{i} D3 ($\lambda$ 5876 \AA) lines.

We analyse the Balmer line variability and phase behaviour. We confirm these lines as highly variable, with excess emission that originates from extended structures and exhibits clear signs of rotational modulation. We have described the line profiles for two distinct states of activity, using different modelling approaches. Similar techniques were applied to the \ion{He}{i} D3 spectra.

A large flare event lasting several days was detected in both Balmer lines and in the \ion{He}{i} D3 line. The energy released during this flare in H$\alpha$\ is of the order of 10$^{37}$~erg, making it the largest H$\alpha$\ flare reported on a cool star.

Our results confirm the extreme complexity of the circumstellar environment of FK Comae. The activity level of this star is quite variable demanding different approaches to the line profile analysis. 


\end{abstract}

\section{Introduction}

FK Comae Berenices (\object{HD~117555}), with its extreme activity, is a key target to study in the solar-stellar connection context. It is a rapidly rotating and apparently single G5 II giant, even though the spectral classification can go as late as G8~III, due to the fast rotation rate and spectral oddities. This star has a rotational velocity of {\it v~$\sin$~i}~=~162.5~$\pm$~3.5~km~s$^{-1}$ and radial velocity -24~$\pm$~3~km~s$^{-1}$ (Huenemoerder et al. \cite{huenemoerder} hereafter refered to as HRBN). The photometric and rotational period is 2.400 day (Jetsu et al. \cite{jetsu}). This gives a projected stellar radius of 7.7~$\pm$~0.2~R$_{\sun}$ (HRBN). As a member of the \object{HR 1614} old disk aggregate, its distance and mass are approximately known: d~$\sim$~200--300~pc and M$~\sim$~1.5~M$_{\sun}$ (Eggen \& Iben \cite{eggen}).

Some spectral peculiarities of FK Com were first observed by Merrill (\cite{merrill}). Amongst the oddities present in the spectra of this star 
can be included a strong, broad and erratically variable H$\alpha$\ emission (e.g. Ramsey et al. \cite{ramsey}), strong UV chromospheric emission lines reminiscent of those observed in RS CVn binaries (Bopp \& Stencel \cite{bopp}), a quasi-sinusoidal light curve (Holtzman \& Nations \cite{holtzman}) and high X-ray luminosity (Walter \cite{walter}).
Some of these spectral characteristics led Bopp \& Stencel (\cite{bopp}) to define a new class of variable stars with FK Comae as the prototype. The FK Comae-type variables were thus defined as rapidly rotating, single, G-K type giants with strong chromospheric and transition region emission. Stars belonging to this class are extremely rare. The space density of these stars was estimated by Collier (\cite{collier}) as 2~$\times$10$^{-8}$~pc$^{-3}$, indicating, when compared with the space density of the rare Hertzsprung-gap G giants, that the evolution through this part of the H-R diagram must be extremely rapid.

Typical values of rotation rate from G-type stars are $<$~10~km~s$^{-1}$ for single stars and 30--50~km~s$^{-1}$ for stars in close binary systems. The extreme rotational velocity, near breakup,  of FK Comae raises interesting questions on the possible evolutionary status of this star. Two scenarios were advanced to explain the observed rotational and activity characteristics: accretion from a very small unseen companion or product of the recent coalescence of a close binary system. Based on radial velocity variations measurements, McCarthy \& Ramsey (\cite{mccarthy}) proved the first hypothesis very unlikely, by imposing severe limits to the mass of this unseen companion. The currently accepted theoretical scenario is that FK Comae-type stars are recently coalesced binary systems. According to Webbink (\cite{webbink}), W UMa systems, due to the thermal instabilities caused by the primary's departure from Main-Sequence, evolve into configurations that can only be stabilized by a big loss of angular momentum, coalescing into a single star in a way that is mostly independent of their evolution. The rapid rotation would arise from the angular momentum of the former binary and the activity level from a rotation-driven magnetic dynamo (Ramsey et al. \cite{ramsey}). The descendents of these binaries would be indistinguishable from single stars. Welty \& Ramsey (\cite{welty}) presented evidence that conforms with such scenario, based on the analysis of the width and shape of the spectral lines.

In FK Comae, the most striking spectral feature is the H$\alpha$\ emission line. This line profile is extremely broad, asymmetric and presents variations on different time scales. The same characteristics are present in higher Balmer lines, even though appearing less extreme due to the different optical thickness of the lines. The phase modulation in these lines indicates that the emitting extended material is corotating with the star.
Based on the H$\alpha$ to H$\beta$ flux ratios, HRBN have proposed that the emission arises in structures similar to solar prominences. 

We present here a model of the H$\alpha$\ and H$\beta$\ line profiles for a quiet phase of the circumstellar environment, derived from a data set obtained in May 96. The model includes a 3D gaussian spatial distribution of emitting matter, extending to several stellar radii, and an optically thick near-stellar shell of absorbing material. The data sets from May and June 97 represent a more active state, demanding a different modelling technique. The H$\beta$\ line profiles present a complex structure that we model as the superposition of gaussian-shaped components. The same components were found to be present in the H$\alpha$\ line profiles, superposed on a more distributed ``bulk'' emission. We were able to track these components in the velocity vs phase diagram for approximately 3 rotational periods. A large flare event was detected in our May 97 data set, in the Balmer and \ion{He}{i} D3 lines, and we computed the corresponding energy budget.

\section{Observations and data reduction}

The MUlti-SIte COntinuous Spectroscopy (MUSICOS) project was developed with the aim of providing continuous spectroscopic coverage (in a large wavelength range and with instruments as similar as possible) for a wide range of stellar objects (e.g. Foing \& Catala \cite{foingb}; Foing et al. \cite{foingd}; Catala et al. \cite{catalaa}, \cite{catalab}; Huang et al. \cite{huang}). Following these orientations, the first fiber-fed echelle cross-dispersed MUSICOS spectrograph was designed and built in Meudon-Paris Observatory (Baudrand \&  B\"{o}hm \cite{baudrand}) and a replica was built at the ESA Space Science Department in ESTEC and adapted for the INT. The ESA-MUSICOS spectrograph was installed at the 2.5~m Isaac Newton Telescope (INT), at the {\em Observatorio del Roque de los Muchachos}, La Palma, Spain.

The commissioning campaign of this instrument took place from April 29th to May 6th. Our first data set was obtained in this campaign. The resolving power achieved was $\lambda/\Delta\lambda~\sim$~35\,000. The second data set was obtained in May~97 also at the INT with the same instrument.  The CCD detector used in both runs was a Tektronix, 1024~$\times$~1024 pixels, with a readout noise of $\sim$~5~e$^{-}$. To reduce the INT data, we used the MIDAS Nov.~94 and Nov.~96 echelle reduction packages. The reduction procedure used was standard for echelle spectra, with some problems in the correction for the blaze function due to the low level of counts of the flat-field exposures in the bluer part of the spectra.

In addition to these INT observations, we obtained in June~97 more H$\alpha$\ spectra at the {\em Observatoire de Haute Provence} (OHP), in France. We used the Aur\'{e}lie spectrograph at the 1.52~m coud\'{e} telescope, with resolving power of 30\,000. The reduction of this data set was done using the on-line reduction facilities available on site.

In Table~1 we give the summary information on all the spectra obtained. The ephemeris used for the phase computations is from Chugainov (\cite{chugainov}): 2\,442\,192.345~+~2.400E.

\begin{table}[ht]
\begin{center}
\caption{List of 1996 and 1997 FK Comae observations}
\begin{tabular}{clcrlr}
\hline
HJD$^{*}$&\hspace{0.3cm}Date&UT$^{*}$&t$_{e}^{**}$&site&$\phi^{***}$\\
\hline
2\,450\,205.38&96~May~1&21:08:20&\llap{1}800&INT&.764\\
2\,450\,205.41&96~May~1&21:45:40&\llap{1}800&INT&.775\\
2\,450\,205.44&96~May~1&22:38:41&\llap{1}800&INT&.790\\
2\,450\,205.46&96~May~1&23:09:12&\llap{1}800&INT&.800\\
2\,450\,583.47&97~May~14&23:20:57&900&INT&\llap{0}.343\\
2\,450\,583.48&97~May~14&23:39:03&\llap{1}800&INT&\llap{0}.345\\
2\,450\,584.36&97~May~15&20:37:44&\llap{1}500&INT&\llap{0}.712\\
2\,450\,584.50&97~May~16&00.05:44&800&INT&\llap{0}.772\\
2\,450\,585.37&97~May~16&20:49:30&900&INT&\llap{1}.132\\
2\,450\,585.39&97~May~16&21:06:29&\llap{1}200&INT&\llap{1}.137\\
2\,450\,585.39&97~May~16&21:27:55&\llap{1}200&INT&\llap{1}.143\\
2\,450\,585.41&97~May~16&21:48:20&\llap{1}200&INT&\llap{1}.149\\
2\,450\,587.38&97~May~18&21:08:37&\llap{1}200&INT&\llap{1}.971\\
2\,450\,587.40&97~May~18&21:29:41&\llap{1}200&INT&\llap{1}.977\\
2\,450\,588.38&97~May~19&21:06:12&\llap{1}200&INT&\llap{2}.387\\
2\,450\,588.39&97~May~19&21:26:56&\llap{1}200&INT&\llap{2}.393\\
2\,450\,589.38&97~May~20&21:08:28&\llap{1}200&INT&\llap{2}.804\\
2\,450\,589.39&97~May~20&21:29:14&\llap{1}800&INT&\llap{2}.810\\
2\,450\,590.38&97~May~21&21:08:30&\llap{1}800&INT&\llap{3}.221\\
2\,450\,590.40&97~May~21&21:38:54&\llap{1}200&INT&\llap{3}.230\\
2\,450\,617.38&97~June~17&21:04:56&\llap{1}200&OHP&\llap{14}.470\\
2\,450\,617.43&97~June~17&22:12:53&\llap{1}200&OHP&\llap{14}.489\\
2\,450\,617.47&97~June~17&23:16:24&\llap{1}200&OHP&\llap{14}.508\\
2\,450\,623.39&97~June~23&21:20:55&\llap{1}200&OHP&\llap{16}.974\\
2\,450\,623.41&97~June~23&21:44:47&\llap{1}200&OHP&\llap{16}.981\\
2\,450\,623.42&97~June~23&22:08:28&\llap{1}200&OHP&\llap{16}.988\\
2\,450\,623.44&97~June~23&22:33:26&\llap{1}200&OHP&\llap{16}.996\\
2\,450\,623.46&97~June~23&22:58:06&\llap{1}200&OHP&\llap{17}.003\\
2\,450\,623.48&97~June~23&23:22:20&\llap{1}200&OHP&\llap{17}.010\\
\hline
\end{tabular}
\end{center}
$^{*}$ Times given at the start of the exposures.\\
$^{**}$ Exposure time in seconds.\\
$^{***}$ A rotation-number ordinate was added to the $\phi$\ value of the spectra of 1997 to represent the different rotation periods covered.
\end{table}

\subsection{Artificial template for excess spectra}

As we are mainly interested in the study of phenomena related with the excess activity, one more standard reduction step was applied to all the spectra. As long as the line profile has the same shape over the stellar disc, a rotationally broadened flux profile can be obtained by convolving the flux profile of a non-rotating template star with the rotational profile of the observed star. As a template, we used \object{$\kappa$~Hercules} (\object{HD~145001}), a G8~III type star, still adequate for differential comparison. We have artificially broadened the spectrum of this low activity star to simulate the rotational broadening of FK~Com. This reproduces adequately the photospheric lines and serves as baseline for the chromospheric lines. After normalization, these model spectra were subtracted from the ones of FK~Com, for the spectral orders of interest. 

\section{Results from May~96 line profiles}

\subsection{Low activity Balmer extended emission}

\begin{figure}[ht]
\resizebox{\hsize}{!}{\includegraphics{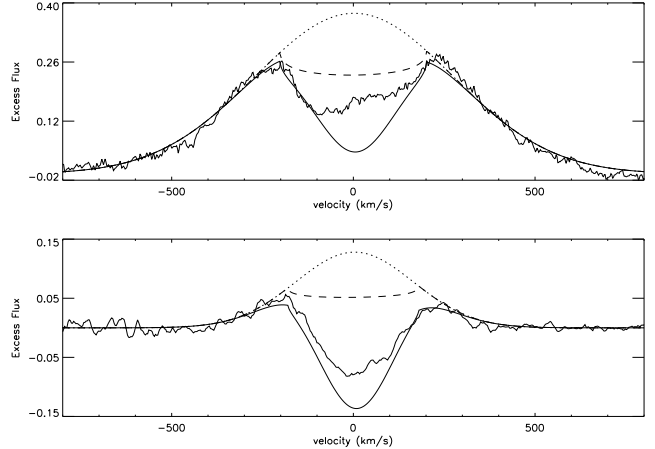}}
\caption{The H$\alpha$ (top) and H$\beta$ (bottom) average excess spectra on 1st May~96, as well as the gaussian profiles (dotted line), fitting the wings of the extended emission. The obscuration by the stellar disc and the absorption of the near-stellar shell are displayed (dashed line) and also the effect of the extra absorption of the background stellar profile (full line).}
\label{model}
\end{figure}

From 1st May~96, we describe 4 FK~Com spectra of H$\alpha$\ and H$\beta$. These spectra are symmetric and the measured equivalent widths (respectively -2.4 and 1.0 \AA) are low when compared to other available measurements (both from the literature and from our measurements). This suggests that these spectra represent a low activity state for this star.

The difference spectra show emission extending up to $\sim$~780~km~s$^{-1}$ in H$\alpha$ and $\sim$~450~km~s$^{-1}$ in H$\beta$. We found that in both lines the wings of the extended emission, outside {\it v~$\sin$~i}, can be fitted by gaussian profiles, as it can be seen in Fig.~\ref{model}. The full width at half maximum of the fitted gaussian profiles are 642~$\pm$~12~km~s$^{-1}$ and 381~$\pm$~20~km~s$^{-1}$ respectively for H$\alpha$\ and H$\beta$. Assuming that the velocities correspond to corotating material, we derive the maximum extent of the emission, $\sim$~4.75~{\it R$_{*}~\sin$~i} for H$\alpha$ and $\sim$~2.66~{\it R$_{*}~\sin$~i} for H$\beta$ ({\it R$_{*}~\sin$~i} is the projected stellar radius). 

Using the Cousins-Bessel magnitudes for FK Com, m$_{R}$~=~7.66 and m$_{B}$~=~9.07 (Holtzman \& Nations 1984) and the corresponding flux calibrations (Bessel 1979) the Balmer line fluxes were computed. The ratio between the fluxes of the H$\alpha$ and H$\beta$ for the fitted gaussian is at maximum emission $\sim$~4, and $\sim$~7 for the integrated flux, reminiscent of solar-like active prominences (Oliveira et al. \cite{oliveiraa}). However, as we derived different widths for H$\alpha$\ and H$\beta$ indicating a different extension of the emitting plasmas, these global ratios may have little direct physical meaning.

\subsection{Modelling of circumstellar emission}

We have modelled the emission wings by assuming an effectively thin, corotating emission and a 3D gaussian distribution of the source function. Under such assumptions, the measured full width relates directly to the scale height of the emission. We have performed numerical integration of such a 3D distribution at each velocity. This characterization represents well the observed wings, but differs strongly in the central absorption profile, within $\sim~\pm$~1.3~{\it v~$\sin$~i} from the rotational axis.

The central absorption, superposed to the emission, can partly be interpreted as the presence of the star itself and the effect of obscuration from the stellar disc. The analysis of the profiles shows that this absorption is in fact slightly broader than 2~$\times$~{\it v~$\sin$~i}, in the two Balmer lines, suggesting the presence of a near-stellar shell of absorbing material.
The radius of this absorption shell R$_{abs}$ is one of the parameters to be adjusted in this modelling approach. The presence of this shell also contributes with another extra absorption, as it absorbs the background stellar profile by a factor $\varepsilon$. 

The numerical integration of the 3D gaussian emission is computed at a given velocity, removing the volume obscured by the star and the absorption shell. The modelled profile obtained by this method is then corrected for the extra absorption, by subtracting the function $\varepsilon~\times$~I$_{b}$, where I$_{b}$ is the background stellar spectrum. 

\begin{figure*}[ht]
\resizebox{\hsize}{18.3cm}{\includegraphics{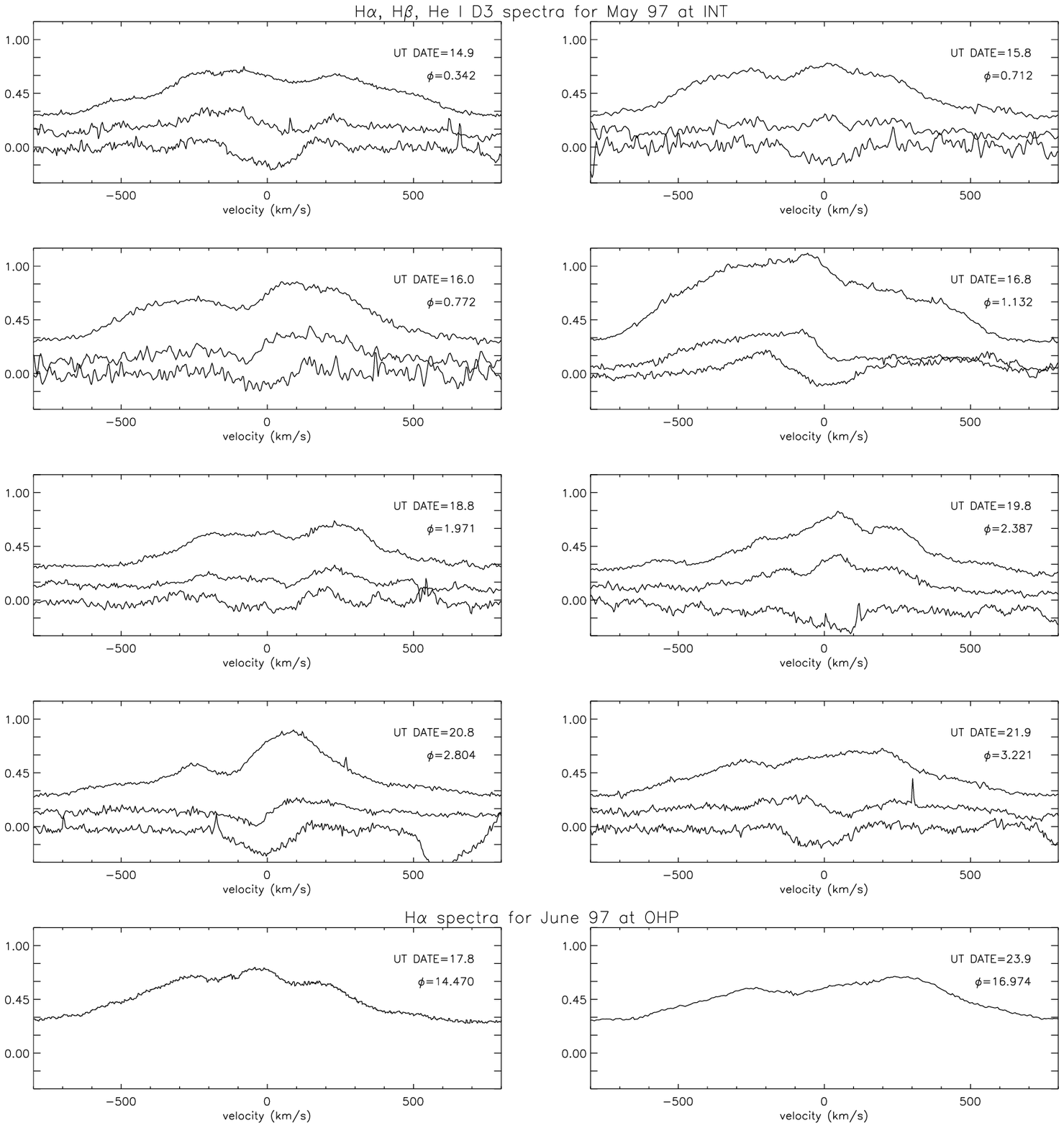}}
\caption{FK Comae excess spectra from May~97 at INT (upper four rows in the graph) and June~97 at OHP (bottom row). In the upper graphs (INT), we plot the H$\alpha$, H$\beta$\ and \ion{He}{i}~D3 spectra. The \ion{He}{i}~D3 spectra were multiplied by a factor 2, to improve features visibility. The H$\alpha$\ and H$\beta$\ spectra were displaced vertically by 0.3 and 0.15 respectively for clarity. In the lower row we show the H$\alpha$\ spectra obtained at OHP, using the same plotting scale. In each graph the phase and UT~date are also indicated. The ephemeris used is from Chugainov (\cite{chugainov}): 2\,442\,192.345~+~2.400E. The integer rotation-number ordinate describes the continuity of the data set, with the first spectrum of 14~May as reference. Velocity components are noticeable in the spectra and their variations in velocity and intensity were followed.}
\label{all97}
\end{figure*}

The resulting profile is compared with the difference spectrum in Fig.\ref{model}. The values for the parameters that give the best agreement with the observed line profiles are R~$_{abs}^{H\alpha}$~=~1.25~{\it R$_{*}~\sin$~i}, R$_{abs}^{H\beta}$~=~1.1~{\it R$_{*}~\sin$~i} and $\varepsilon$~=~0.225.

The fractional depression of the line profiles is approximately the same in these two Balmer lines, indicating possibly an optically thick absorption. Under this assumption, the value of $\varepsilon$ can give us an estimation of the fraction of the stellar disc obscured by the near-stellar shell; according to this model we have a disc coverage of about 23~\%. 

The residuals, after this axisymmetric model contribution is removed, have an H$\alpha$ to H$\beta$ flux ratio of the order of 4, indicating an optically thicker source than the extended emission. Their projected velocities are within  $\pm~v~\sin~i$. Thus, they can be interpreted as plage-like structures and they account for $\sim$~10~\% of the total emission.

\section{Results from May/June 97 line profiles}

\subsection{Multi-component analysis}

In May~97, FK~Comae spectra of H$\alpha$, H$\beta$\ and \ion{He}{i}~D3 were obtained at the INT. This data set is shown in the 4 upper rows of graphs in  Fig.~\ref{all97}. Also in Fig.~\ref{all97}, in the lower row we show the H$\alpha$\ spectra obtained at the OHP, 3 in June~17th and 6 in June~23rd. Each pannel shows the measured spectral lines for representative times, after removing the rotationally broadened template spectra (each spectrum is an average of the spectra taken in each night).

We have tried to model these spectral lines in a similar way as it was done for the May~96 data set. Clearly, that axisymmetric model does not apply to these May/June~97 Balmer line profiles, very asymmetric, variable and with a more complex component structure. Thus, we used a multi-component approach, that consists in fitting a variable number of discrete gaussian components to each of the spectral lines, representing emitting structures. We also analysed the phase behaviour of these structures. We plot these components in velocity vs phase ($\phi$). We fit the velocity curves that describe the phase behaviour of the emitting structures as they corotate with the star, by an equation of the type:
$$\frac{R}{R_{*}}~\times~{\it v_{eq}~\sin i}~\cos \theta~\sin(\phi~+~\phi_{0})~=~v_{c}~\times~\sin(\phi~+~\phi_{0})$$ 
where {\it v$_{eq}$} is the equatorial rotational velocity, $\theta$ is the stellar latitude, $\phi_{0}$ the stellar longitude and {\it v$_{c}$} is the velocity excursion in km~s$^{-1}$. The two Balmer lines were analysed in parallel to investigate their similarities. The same multi-component approach was also used for the \ion{He}{i} D3 line.

\subsection{The Balmer lines fitted components}

\begin{figure}[ht]
\resizebox{\hsize}{!}{\includegraphics{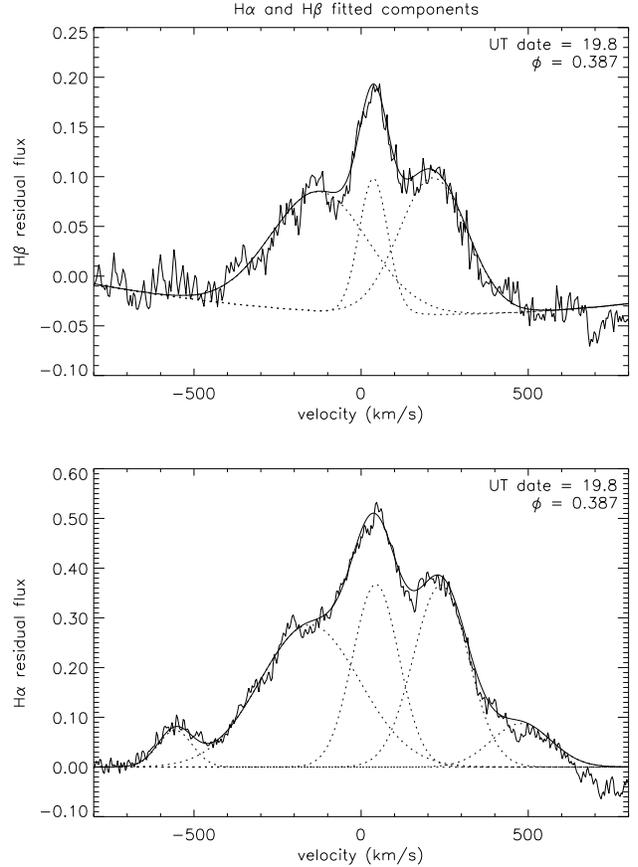}}
\caption{An example of the multi-component fit to the Balmer line profiles (from top to bottom H$\beta$\ and H$\alpha$). It is clear that the same components are present in both spectral lines, except for the higher velocity ones, that are masked in the noise in H$\beta$.}
\label{components}
\end{figure}

Figure~\ref{components} shows an example of how the profiles were fitted with gaussian emission components. Comparing the two Balmer line profiles in Figs.~\ref{all97} and \ref{components} it is clear that the same components seem to be present in both spectra. This becomes even more evident when the average H$\alpha$\ and H$\beta$\ profiles are subtracted from the individual profiles. The analysis of these average profiles hints that the H$\beta$\ emission is somehow dominated by discrete components, while in H$\alpha$\ these same components are masked by a more stationary, distributed, ``bulk'' emission.

\begin{table*}[ht]
\begin{center}
\begin{tabular}{ccccccccc}
\hline
Date~97&$\phi$&v$_{\alpha}$/{\it v~$\sin$~i} &(FWHM)$_{\alpha}$/{\it v~$\sin$~i}&A$_{\alpha}$&v$_{\beta}$/{\it v~$\sin$~i} &(FWHM)$_{\beta}$/{\it v~$\sin$~i} &A$_{\beta}$&identification\\
\hline
14~May&0.343&\llap{-}3.3&1.0&0.08&*&*&*&\\
&0.343&\llap{-}1.2&2.4&0.35&\llap{-}1.1&1.8&0.13&A\\
&0.343&1.5&3.2&0.34&1.8&1.7&0.06&B\\
15~May&0.712&\llap{-}1.8&2.6&0.40&\llap{-}1.6&1.6&0.06&B\\
&0.712&\llap{-}0.1&1.0&0.19&0.0&0.6&0.10&\\
&0.712&1.1&2.1&0.40&1.2&1.2&0.08&A\\
&0.712&3.4&1.1&0.07&*&*&*&\\ 
15~May&0.772&\llap{-}2.2&2.2&0.29&\llap{-}1.7&1.4&0.06&B\\
&0.772&0.6&3.3&0.53&0.9&1.8&0.16&A\\
16~May&0.132&\llap{-}1.9&2.9&0.67&\llap{-}1.5&2.5&0.21&A\\
&0.132&\llap{-}0.3&1.2&0.31&\llap{-}0.4&0.6&0.11&\\
&0.132&1.4&3.0&0.47&1.9&2.5&0.09&B\\
18~May&0.971&\llap{-}0.6&2.9&0.32&\llap{-}0.9&1.6&0.06&C\\
&0.971&1.5&1.3&0.25&1.4&1.0&0.10&\\
&0.971&2.3&2.9&0.10&2.9&0.3& 0.04&\\
19~May&0.387&\llap{-}3.4&0.7&0.07&*&*&*&\\ 
&0.387&\llap{-}0.9&2.2&0.29&\llap{-}0.8&2.1&0.12&\\
&0.387&0.3&1.0&0.37&0.2&0.6&0.14&A\\
&0.387&1.5&1.2&0.37&1.3&1.5&0.14&B\\
&0.387&2.9&1.3&0.09&*&*&*&\\
20~May&0.804&\llap{-}2.8&1.9&0.11&\llap{-}2.8&1.6&0.02&C\\
&0.804&\llap{-}1.5&0.9&0.22&*&*&*&B\\
&0.804&0.4&1.8&0.50&1.0&1.3&0.08&A\\
&0.804&1.9&3.4&0.12&*&*&*&\\
21~May&0.221&\llap{-}1.7&2.6&0.26&\llap{-}0.7&1.2&0.08&\\
&0.221&0.9&2.5&0.38&1.6&1.0&0.07&B\\
&0.221&3.1&1.3&0.07&2.9&1.0&0.03&C\\
17~June&0.470&\llap{-}3.4&0.4&0.02&**&**&**&\\ 
&0.470&\llap{-}1.1&3.5&0.43&**&**&**&\\ 
&0.470&\llap{-}0.2&0.7&0.16&**&**&**&\\
&0.470&1.3&1.4&0.20&**&**&**&\\
&0.470&2.9&1.3&0.07&**&**&**&\\
23~June&0.988&\llap{-}1.7&2.8&0.27&**&**&**&\\
&0.988&0.2&0.5&0.04&**&**&**&\\
&0.988&1.6&3.0&0.39&**&**&**&\\
\hline
\end{tabular}
\end{center}
\caption{Parameters for the fitted H$\alpha$\ and H$\beta$\ components for the INT May~97 and OHP June~97 campaigns. v$_{\alpha}$ and (FWHM)$_{\alpha}$ are respectively the velocity position of the maximum and the full width at half maximum (in units of {\it v~$\sin$~i}~=~162.5~km~s$^{-1}$) for each of the fitted gaussian components in the H$\alpha$\ excess spectra. The velocity errors are typically 0.1--0.3 {\it v~$\sin$~i}. A$_\alpha$\ is the amplitude (in continuum units) for each of the components. v$_{\beta}$, (FWHM)$_{\beta}$ and A$_{\beta}$ are the corresponding quantities for the H$\beta$\ components. Some H$\alpha$\ components do not have measurable H$\beta$\ counterparts (indicated by *). For the June~97 data set (**), no H$\beta$\ observations were made. The capitals A, B and C in the last column identify some of the components which behaviour is described by the velocity curves plotted in Figs.~4 and 5.}
\end{table*}

In Table~2 we give the parameters for each of the gaussian components fitted to the Balmer line profiles, H$\alpha$\ and H$\beta$. In Fig.~\ref{vel_all}, we show several plots that describe their behaviour: from top to bottom, the peak intensity vs velocity, and velocity vs phase for both lines. In order to highlight their phase behaviour, we also plotted (in the two lower graphs) relevant velocity curves. Figure~\ref{vel_time} shows the velocity evolution of the components in time. We have estimated a maximum error in the velocity determinations of about 50~km~s$^{-1}$. The plots for these two Balmer lines were analysed together in order to clarify the structures' position and evolution. 

\subsubsection{Velocity curves analysis}

\begin{figure*}[ht]
\resizebox{\hsize}{13cm}{\includegraphics{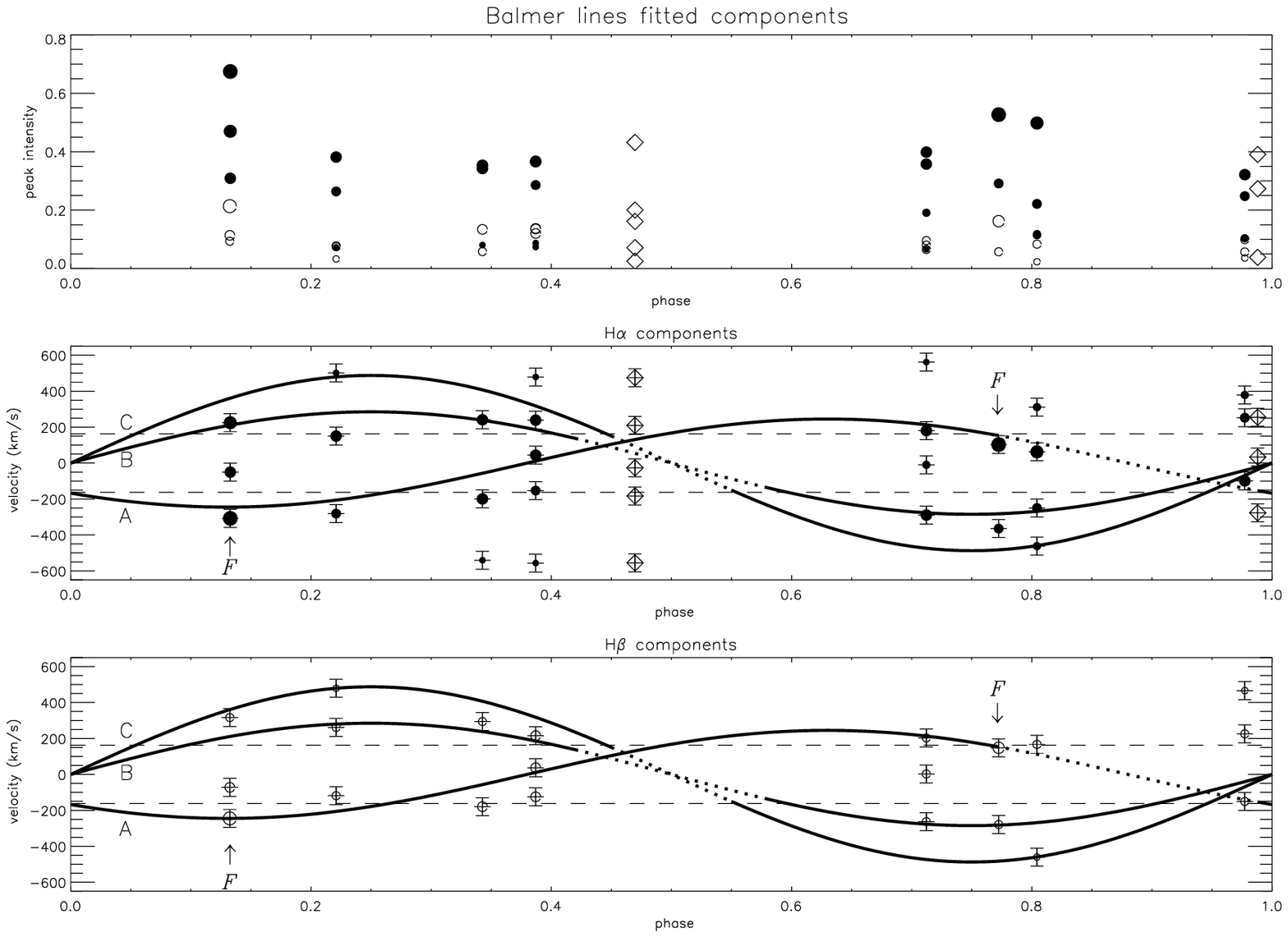}}
\caption{Phase behaviour of Balmer line components. From top to bottom: peak intensity plotted against phase with the size of the symbols representing the corresponding intensity, $\bullet$ represents H$\alpha$ and $\circ$ H$\beta$\ from INT May~97 and $\diamond$ H$\alpha$ from the OHP June~97; velocity evolution of the H$\alpha$\ components with phase; velocity evolution of the H$\beta$\ components with phase. In the bottom two graphs, three velocity curves (A, B and C) were overplotted, corresponding to different distances from the rotational axis and longitudes: v$_{A}$~=~245~$\times~\sin(\phi~+~0.62)$, v$_{B}$~=~285~$\times~\sin(\phi)$ and v$_{C}$~=~487~$\times~\sin(\phi)$. The dotted part of these curves represent the occultation of the structures by the star. The points indicated with ``{\it F}'' are flare spectra. The dashed lines represent $\pm$~{\it v~$\sin$~i} from the rotational axis.}
\label{vel_all}
\end{figure*}

As it can be seen in Fig.~\ref{vel_all}, three structures could be followed in phase, corresponding to v$_{A}$~=~245~$\times\sin(\phi+0.62)$, v$_{B}$~=~285~$\times\sin(\phi)$ and v$_{C}$~=~487~$\times\sin(\phi)$, both in H$\alpha$\ and H$\beta$. It was observed that most of the components were detected within approximately $\pm$~2~{\it v~$\sin$~i} from the rotational axis, in agreement with Welty et al. (\cite{welty et al}) that placed the emitting material at $\leq$~2~{\it R$_{*}$}.

Similar component analysis was carried out with the H$\alpha$\ spectra obtained at the OHP in June~97. Two representative spectra were fitted and the velocity positions of the components were compared with what was obtained for May~97. Even though some profiles appear similar (Fig.~\ref{all97}, for phase 1.971 and 16.974), the previously fitted velocity curves do not seem to describe the phase behaviour now observed (Fig.~\ref{vel_all} middle graph). In particular, one can notice the appearance of a new component near-zero velocity at phase $\sim$~0.47, compared with one month earlier. However, the poor phase coverage in June~97 makes a detailed comparison ambiguous. We also find an increase of H$\alpha$ equivalent width of 2~\AA\ in June~97 compared to the quiet level of May~97, for the two phases shown. 

\subsubsection{Comparison of H$\alpha$\ and H$\beta$ components}

Most of the components detected in H$\alpha$\ have their counterparts in H$\beta$. Still, some higher velocity H$\alpha$\ components are not seen in H$\beta$. This is due to the better detection sensitivity in H$\alpha$ as the signal-to-noise ratio in H$\beta$\ is lower, but also to the presence of the ``bulk'' emission in H$\alpha$\ that extends to higher velocities (as it was also seen in the quiet state in May~96 we modelled earlier). 

A slight mismatch can be observed in the velocity position of the components in both lines. Intrinsic measurement uncertainties and physical phenomena combine to create this effect. The origin of the emission is obviously not in point sources, their vertical extent making difficult to determine the center of mass of the matter distribution, due to projection effects and emission gradients in the extended structures. This lack of exact agreement between the two lines is then justified given the simplified geometrical model adopted and the ignoring of radiative transfer effects.

\begin{figure*}[ht]
\resizebox{\hsize}{10.cm}{\includegraphics{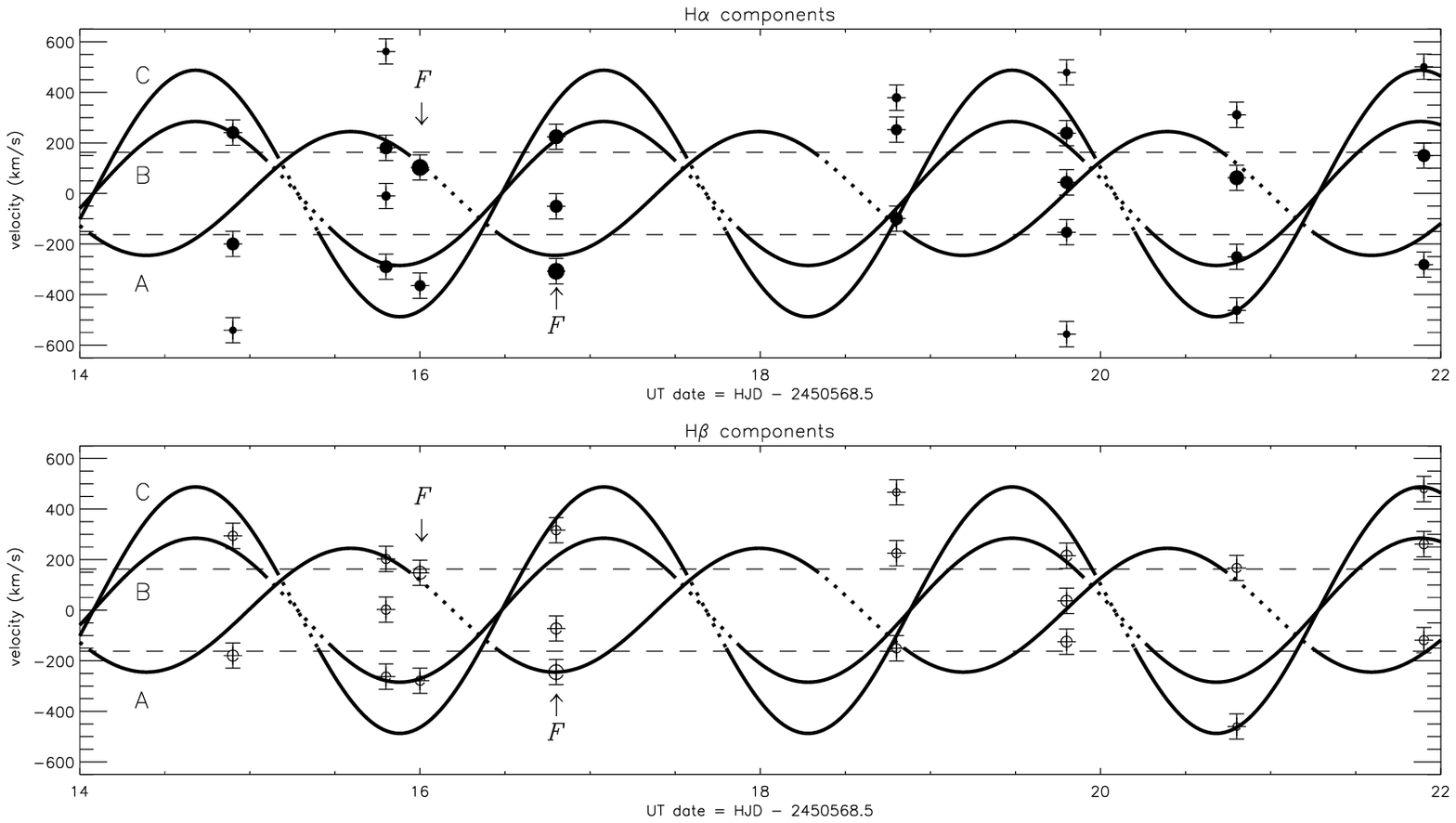}}
\caption{Time behaviour of Balmer line components in May~97 (same plotting code as Fig.~\ref{vel_all}). The velocity curves of the previous figure were also overplotted.}
\label{vel_time}
\end{figure*}

\subsection{The \ion{He}{i} D3 line profiles}

The \ion{He}{i} D3 ($\lambda$ 5876~\AA) line is very weak or absent in inactive stars' spectra. The presence of this line in cool stars like FK Com indicates the existence of non-radiative heating processes, since the line cannot be excited at photospheric temperatures. In FK Comae, this line has two components: an excess absorption and near-surface emission wings. A non-uniform shell of excited \ion{He}{i} could explain this line profile, with the corona as source of ionizing radiation as it is observed in the Sun and active stars.

A similar multi-component study was made for the \ion{He}{i} D3 line. All the spectra present an absorption at near-zero projected velocity, even though its strength is variable. Emission wings were detected above the limb, confirming the existence of a slightly extended shell of excited \ion{He}{i}. The variability observed in the spectra can represent a back-radiation of a nonuniform coronal emission onto the upper chromosphere. The modelling technique we used did not allow us to take advantage of the rotational phase information that might be present in the \ion{He}{i} data set, due to the complexity of the profiles.

\nopagebreak

\subsection{The 16~May~97 long-duration giant flare event}

A large flare event was detected in these two Balmer lines and also in \ion{He}{i} D3. The flaring material seems to extend from the stellar surface up to at least 3 stellar radii.

The two spectra where we detected an increase in the excess emission, what we call a pre-flare spectrum and a flare spectrum, (respectively UT date 16.0 and 16.8 in Fig.~\ref{all97}) are marked by ``{\it F}'' in Figs.~\ref{vel_all} and \ref{vel_time}.

If these spectra represent the same flaring structure viewed at different phases, we can predict a velocity curve allowing to identify the post-flare structures at later phases. Such a velocity curve is also plotted in Fig.~\ref{vel_time}, corresponding to {\it v$_{c}$}~$\sim$~245 kms$^{-1}$ (R~=~1.50~{\it R$_{*}~\sin$~i}) and $\phi_{0}$~=~0.612. At five other phase positions the same structure is identified. Indeed, this structure was first detected on May 15.8 and it was still present on May 21.8, when our run ended.
When transiting in front of the stellar disk, near zero projected velocity, this structure (extended vertically) would have smaller projected dimension, and in the spectrum we should detect a narrow peaked component. This is in fact observed in the spectrum corresponding to UT date 19.8 in Fig.~\ref{components}.

\begin{figure}[t]
\resizebox{\hsize}{7.cm}{\includegraphics{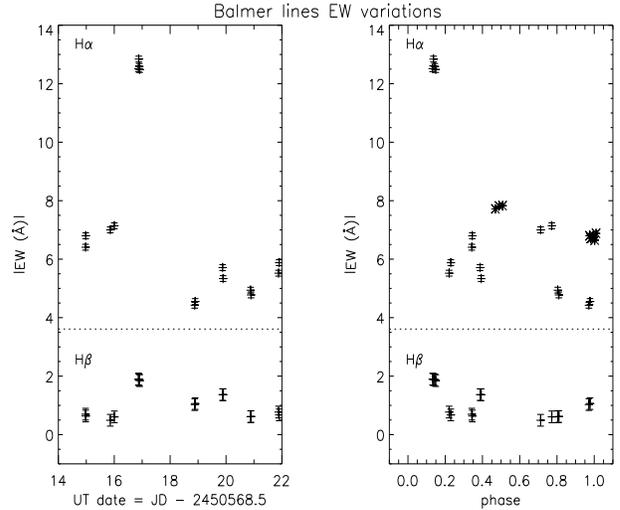}}
\caption{The equivalent width (EW) variations for H$\alpha$\ and H$\beta$, for May/June~97, both in UT date and phase. The lower values in each plot represent H$\beta$. The dotted line indicates the lower H$\alpha$\ EW value measured for the average spectrum from May~96. The symbols $\ast$\ represent the measurements from the OHP. The errorbars are 0.1 and 0.2 respectively for the H$\alpha$\ and H$\beta$\ measurements.}
\label{ew}
\end{figure}

This flare event detected in the Balmer lines caused an increase in equivalent width of 7~$\pm$~1~\AA\ in H$\alpha$\ and 1.2~$\pm$0.2~\AA\ in H$\beta$\ (Fig.~\ref{ew}). In the case of H$\beta$\, the measurements were very difficult, the flare increase in equivalent width being the only variation detected above the errorbars.  Using the Cousins-Bessel magnitudes for FK Com and assuming a distance of 215~pc (HRBN) some flare energetics were computed. The calculated energy released at flare maximum is 8~$\pm$~1~$\times$10$^{31}$~erg~s$^{-1}$ for H$\alpha$\ and 1.1~$\pm$~0.1~$\times$10$^{31}$~erg~s$^{-1}$ for H$\beta$. The H$\alpha$\ to H$\beta$\ flux ratio is about 7, perhaps indicating much lower densities when compared with solar flares. Again the interpretation of such global ratios is dubious in this type of analysis.

The total energy released during the flare for H$\alpha$\ is 11~$\pm$~3~$\times$10$^{36}$~erg. The uncertainties in the intrinsic flare radiative losses for H$\alpha$\ come mainly from the choice of reference quiet spectrum and residual emission from active regions around the flare area. Comparing this energy with the one reported on the \object{HR 1099} giant H$\alpha$\ flare, $\sim$~4~$\times$10$^{35}$~erg (Foing et al. \cite{foingc}), this seems to be possibly the largest H$\alpha$\ flare reported on a cool star.
What is also remarkable is the long duration of this flare event (total equivalent flare duration approximately 2.5~$\pm$~0.5~$\times$10$^{5}$~s).

The ratios of the continuum UBVRI counterparts to H$\alpha$, for similar flare events, are reported to be from 120 (Avrett et al. \cite{avrett}; Machado et al. \cite{machado}; Houdebine \cite{houdebine}) up to 340 (Foing et al. \cite{foingc}). The continuum UBVRI counterpart of the FK Comae flare could therefore amount to losses of 1.3~$\times$10$^{39}$ erg (for a flare ratio UBVRI/H$\alpha$\ of 120). This can be compared to total radiative budgets for the largest reported cool star flares, 1.2~$\times$10$^{38}$~erg on HR~1099 (Foing et al. \cite{foingc}) or 1.8~$\times$10$^{39}$~erg on \object{YY~Men} (Cutispoto et al. \cite{cutispoto}). Thus, we can speculate that we may have observed an exceptionally energetic flare, but we have no independent photometric measurements to quantify the continuum counterpart. 

An enhancement of \ion{He}{i} emission was measured at pre-flare and flare maximum with equivalent width 0.08 and 0.19 \AA\ respectively and at a velocity consistent with H$\alpha$\ and H$\beta$\ components. It was not measurable during the gradual phase, and in particular not in the post-flare spectra of May~19.8. This is similar with \ion{He}{i} measurements in solar and red dwarf flares, which show a more impulsive behaviour of this line, while the Balmer lines show enhancements both in impulsive and the gradual phase (Foing \cite{foinga}). It seems also that the \ion{He}{i} D3 flare maximum was reached before the H$\alpha$\ flare maximum.

The changes observed in the large flare component width might also be associated with large kinetics (turbulence and flows) related with the flare event (Foing \cite{foinga}, Foing et al. \cite{foingc}).

\section{Conclusions}

Once again, the analysis of the line profiles gives us abundant evidence on the complexity of the circumstellar environment of the peculiar giant star FK Comae.

We have modelled empirically the Balmer line profiles for this star. For a lower activity level, the model pictures a 3D gaussian emission to which a near-stellar thick absorption shell is superposed. 

For the more complex structure of the 1997 profiles a different modelling technique is applied.  We have successfully identified and described the phase behaviour of emission components. The H$\beta$\ spectral line was described as the superposition of several (gaussian) emission components. The velocity vs phase diagram allows to trace the same emitting structures as they corotate with the star. In the case of H$\alpha$, the profile is interpreted as a quasi-stationary distributed ``bulk'' emission, to which discrete components, lasting several days, are superposed. The same components were detected in both Balmer line profiles. 

A strong flare event was detected in these two Balmer lines and also in \ion{He}{i} D3. The flaring material seems to extend from the stellar surface up to at least 3 stellar radii. The evolution of the flaring structure was described by a velocity curve corresponding to {\it v$_{c}$}~$\sim$~245 kms$^{-1}$  or R~=~1.50~{\it R$_{*}~\sin$~i} from the rotational axis. The total energy released during the flare in H$\alpha$\ is $\sim$~10$^{37}$ erg, the largest H$\alpha$\ flare energy reported on a cool star. 

The data set of June~97, as it presents longer time intervals continuously covered, provides some more information. Short time-scale variations were observed, that cannot be associated with rotational modulation. Several variable features are visible; they can be interpreted as emission bursts with a time scale of about 1 hour that can be related to the excitation and recombination of giant magnetic loop structures around the star (Oliveira et al. \cite{oliveirab}).

The \ion{He}{i} D3 line is always present in the spectra, with absorption and emission wings components, as a clear indicator of the activity in this star. An absorption near-zero projected velocity is always present with variable intensity. The profiles also present near-stellar emission wings, particularly enhanced during the described flare event.

The need is clear for a multi-site continuous data set to disentangle the intrinsic variations from rotational modulation. We have organized such a campaign in February/March 1998, involving the following telescopes/observatories: INT (with the ESA-MUSICOS spectrograph), Telescope Bernard Lyot, France (with the MUSICOS spectropolarimeter), David Dunlop Observatory (Canada), Russian Special Astrophysical Observatory (SAO) and McDonald Observatory. The phase coverage thus obtained represents a major improvement in the analysis of the circumstellar emission on FK Comae. This will be reported in a subsequent paper (Oliveira et al. in preparation).

\begin{acknowledgements}

J.M.~Oliveira's research work is being supported by the Praxis XXI grant BD9577/96 from the {\it Funda\c{c}\~{a}o para a Ci\^{e}ncia e a Tecnologia}, Portugal. The authors would like to thank the ING/INT and OHP personnel for their support. We thank P.~Sonnentrucker, P.~Ehrenfreund, J.A.~de~Jong, H.C.~Stempels, R.~Le~Poole, P.~Gondoin, T.~Beaufort, J.~Telting, N.~Walton and J.Th.~van~Loon for support observations and for discussion. We acknowledge the referee L.W.~Ramsey for useful comments. Jacco, you know what? You know...

\end{acknowledgements}

\end{document}